\documentclass[aps,pre,amssymb,twocolumn,showpacs]{revtex4}
\usepackage{graphicx}
\usepackage{epsfig}
\usepackage{subfigure}
\usepackage{graphics}
\usepackage{setspace,bm}
\begin{document}

\renewcommand{\bottomfraction}{0.8}
\setcounter{bottomnumber}{3}

\bibliographystyle{aip}

\title{Infrared actuation in aligned polymer-nanotube composites}
\author{S. V.~Ahir, A.M.~Squires, A.R.~Tajbakhsh and E.M.~Terentjev}
\affiliation{Cavendish Laboratory, University of Cambridge, J.J.
Thomson Avenue, Cambridge CB3 OHE, U.K. }

\begin{abstract}
\noindent Rubber composites containing multi-walled carbon
nanotubes have been irradiated with near infrared light to study
their reversible photo-mechanical actuation response. We
demonstrate that the actuation is reproducible across differing
polymer systems. The response is directly related to the degree of
uniaxial alignment of the nanotubes in the matrix, contracting the
samples along the alignment axis. The actuation stroke depends on
the specific polymer being tested, however, the general response
is universal for all composites tested. We conduct a detailed
study of tube alignment induced by stress and propose a model for
the reversible actuation behavior, based on the orientational
averaging of the local response. The single phenomenological
parameter of this model describes the response of an individual
tube to adsorption of low-energy photons; its experimentally
determined value may suggest some ideas about such a response.
\end{abstract}


\maketitle

\section{Introduction}

Many structures are able to change their mechanical properties and
dimensions when an appropriate stimulus is applied. This
phenomenon is commonly called actuation. The energy from an
external source triggers changes in the internal state of the
system, leading to a mechanical response much larger than the
initial input. This ability to unlock internal work in a solid
state structure is of key importance for many actuator
applications. Actuators with differing characteristics and
mechanisms have been widely adopted by industry to fill a variety
of technological requirements~\cite{Huber1997} with some having a
one-way response, while others providing an equilibrium,
reversible response to the given stimulus. Shape-memory
alloys~\cite{refAlloys} or polymers~\cite{Langer01} are good
examples of such smart actuating systems. However, in most cases a
shape memory system works only in one direction, requiring a reset
after the actuation. Only very few systems can reversibly actuate
and then return back to the equilibrium shape once the stimulus is
removed. So far only liquid crystal elastomers~\cite{Eugenebook}
have proven to be a truly equilibrium reversible actuating system.

A polymer benign to external stimulus can also be made to actuate
when blended with of one or more distinctly different materials to
impart a new physical response leading to the actuation process. A
recent article has demonstrated one such system, based on a common
silicon rubber filled with a low concentration of aligned carbon
nanotubes, actuating in response to infrared
radiation~\cite{Ahir2005}. Apart from actuation itself, the
stimulation of functionalized nanotubes by infrared (IR) radiation
is also proving an effective technique, e.g. in biomedical
applications~\cite{Dai2005}. Clearly, there are rich prospects and
much motivation to understand nanotube action and the actuation
behavior under IR irradiation when they are embedded in a polymer
matrix.

The work presented here focuses on the use of multi-walled carbon
nanotubes (MWCNTs) to impart equilibrium mechanical actuation in
the rubbery matrix. The properties of multi-walled nanotubes has
been well documented for over a
decade~\cite{Ajayan1997,Forro2001,PhysicalNanotubesBook}. Their
behavior in polymer composites is less well understood but some
reviews have recently appeared in the
literature~\cite{Breuer2004,Andrews2004,Harris2004,Thostenson2001,Nalwa2005}.
For mechanical applications, the interface between the tube
surface and the host polymer is of critical importance and most of
the studies have focussed on this aspect. In contrast, the nature
of the active response of nanotubes within a polymeric matrix has
yet to be fully understood. The complex behavior of tubes is often
simplified and analogies are made with aligned rigid rods. It is
unclear whether such analogies are always valid, especially when
the tubes do not necessarily form rigid rods in a polymer matrix
and certainly do not align unless an external field is
present~\cite{Nalwa2005}.

The actuating properties of MWCNTs have recently being elucidated
upon with the possibility of designing nanoelectromechanical
(NEMS) systems~\cite{Cumings2000}. The actuator properties of
individual bending MWCNTs under an applied electric field have
been studied experimentally~\cite{Poncharal1999}. The torsional
actuation behavior of multi-walled tubes has also been
reported~\cite{Willams2002,Fennimore2003}. These works are
important but we note that all these studies focus on individual
tubes and not a collection of tubes, nor their properties within a
continuous elastic matrix. The massive elastic response of
single-walled nanotube bundles, when stimulated by light, was very
effectively demonstrated by Zhang and
Iijima~\cite{Zhang1999ElasticLight}, although little work has
followed from their discovery. They showed the bundles responding
to visible light and a near IR laser radiation by elastically
changing their dimensions; examining the figures in
\cite{Zhang1999ElasticLight} we deduce that the induced strain
must be about 20\%. In the context of this paper, we shall refer
to the actuation stroke as the change in strain, when an external
stimulus is applied.

There are several reports of actuation behavior of
polymer-nanotube
composites~\cite{Landi2002,Koerner2004,Tahhan2003}. These works
have focussed on accentuating the already present features of the
host matrix by adding nanotubes. The tubes act to exaggerate the
response by either improving electromechanical properties or
increasing heat transfer efficiency due to the inherent high
conductivity~\cite{Naciri2003,{Nalwa2005}} that originates from
their delocalized $\pi$-bonded skeleton. Recent work has departed
from this traditional `improvement' scheme and asked whether it is
possible to blend nanotubes with benign polymers to create new
composite actuator properties, that otherwise would not occur in
that system. Such effects have been observed by Courty~\textit{et
al.}~\cite{Courty2003} where electric field stimulation of liquid
crystal elastomers with embedded MWCNTs lead to mechanical
contraction.

Similarly, the photomechanical response from MWCNTs when embedded
in a silicone rubber (PDMS) matrix~\cite{Ahir2005} is a new
effect. The pristine elastomer shows no response to near IR
radiation, yet the presence of nanotubes causes a strong
reversible response that can be tailored by manipulating the
degree of alignment the tubes experience. The present work expands
on such a simple polymer nanocomposite system and goes on to show
that the effect can exist independently of the host polymer matrix
which, by the presence of MWCNTs, produces a mechanical response
to the IR irradiation. We show that both a compression and an
extension response can be achieved (depending on the external
uniaxial strain applied to the composite sample), but that the
magnitude of the actuation stroke strongly depends on the host
polymer used. We also develop a simple model that considers the
orientational ordering of nanotubes in the matrix along with their
individual and bulk actuating behavior.

This paper is organized as following: after giving details of
preparation and basic composite characterization, we concentrate
on the analysis of tube orientation induced by stretching of the
host polymer matrix, section III.  We then turn to the
IR-stimulated actuation, section IV, and study different
nanocomposite systems in some detail (although the majority of our
studies remain on the PDMS system). Section V presents a simple
theoretical model that might well describe the actuation mechanism
and compares it with our experimental data and the literature. We
conclude that two-way actuation behavior is dependent on nanotube
orientation, but is independent of the chosen homogenous polymer
matrix and can occur in any rubbery solid, albeit with varying
magnitude. It is thought that no other materials of any class
(metal, polymer, ceramic) can display this behavior and to such
large effect, thus, the study of the underlying physics of such
systems is of clear scientific, medical and commercial importance.

\section{Experimental}
\subsection{Materials}

There are many different sources of carbon nanotubes on the market
today. After extensive searching and testing, we have settled on
nanotubes provided by Nanostructured \& Amorphous Materials, Inc.
(USA). These are multi-walled, with the core diameter between
5-10nm, outer diameter of 60-100nm and length between 5-15
microns. Purity has been verified (with SEM) as $>$95\% in raw
form from the supplier, in agreement with specification. These
nanotubes were not surface-modified at any time during processing
and are used throughout this study for all polymers tested.
Chemical functionalization is necessary in many nanocomposite
fields, but in our work it has been avoided to reduce the number
of variables in the system. We share the views of other authors
that chemical functionalization of the tube walls will degrade the
properties of the tubes overall due to further introduction of
sp$^3$ hybridized carbon defects~\cite{Garg1998,Sinnott2002}.

Three types of polymer have been tested; PDMS rubber (crosslinked
polydimethylsiloxane), SIS (styrene-isoprene-styrene) triblock
thermoplastic elastomer and a nematic liquid crystal elastomer
(LCE, in both mono and polydomain form). Each type of polymer has
a unique preparation method outlined in the following sections.
Where possible, similarities in processing have been kept.
Table~\ref{tableMats} lists the composites made and their
abbreviations.
\begin{table}[b]
\caption{List of host polymer materials, nanotube loading and the
abbreviations of resulting composites.}
 \label{tableMats}
\begin{center}
    \begin{tabular}{|c|p{0.65in}|p{1.3in}|}
    \hline
    \textbf{Host} & \textbf{Tube loading (wt\%)} & \textbf{Abbreviation} \\
    \hline\hline
    PDMS & 0, 0.02, 0.3, 0.5, 1, 2, 3, 4 \& 7 & PDMS, PDMS0.02,
    PDMS0.3, PDMS0.5, PDMS1 ..  \& PDMS7 \\
    \hline
    Mono LCE & 0 \& 0.2 & MLCE \& MLCE0.2 \\
    \hline
    Poly LCE & 0 \& 0.15 & PLCE \& PLCE0.15 \\
    \hline
    SIS & 0.01 & SIS0.01\\
    \hline
\end{tabular}
\end{center}
\end{table}

\subsubsection{PDMS composite preparation}

The PDMS (Sylgard 184$^{TM}$) silicone elastomer system was
obtained from Dow Corning, USA, in the form of the main compound
and the hydrosilane curing agent (crosslinker). In pristine
conditions, the mixing and crosslinking procedure gives a uniform
solvent-free elastomer. We have verified (with SEM on
cryo-microtomed and freeze-fractured surfaces) that the resulting
polymer network is pure crosslinked PDMS with no filler particles,
as sometimes is the case with supplied elastomer mixes.

The nanotube-polymer composite was fabricated by first carefully
weighing the desired quantity of nanotubes and the polymer
compound. Calculations of weight percentage take into account the
weight of crosslinker, to be later used in the mixture. The highly
viscous fluid was sheared using an Ika Labortechnik mixer for a
minimum of 24 hours.

The crosslinker was added to the mixture after 24 hours. The ratio
of crosslinker to PDMS was 1:10, according to Sylgard 184$^{TM}$
specification, ensuring negligible sol fraction after preparation
of the pristine network. The sample was then further sheared for
another 30 seconds before being placed in vacuum for 5 minutes to
degas, at all times remaining at ambient temperature to ensure
little crosslinking reaction takes place in this time. After
removing the air cavities, that unavoidably form during shear
mixing, the mixture was deposited in a specially designed reactor
(centrifuge compartment with PTFE film lining its inner wall) and
placed in a centrifuge at 5000rpm and 80$^{\circ}$C.  At this
temperature the PDMS crosslinking is much faster and the
centrifugation achieves the uniform thickness and full
homogenization of resulting rubber composite samples.

The subsequent processing depends on the target sample properties.
If we require a completely non-aligned nanotube dispersion, the
sample remains in the reactor for 24 hours, resulting in a
homogeneous elastomer composite. In some cases (as will be clear
from the text below) we aim to produce a sample with nanotubes
permanently pre-aligned. In this case the initial mix remains in
the reactor, at 80$^{\circ}$C, for 14 minutes (calculated from
separate measurements of crosslinking reaction rates). The
partially crosslinked network was then removed from the reactor
and aligned mechanically by applying uniaxial extension using
specially designed clamps. Removing the sample from the reactor
after what is a relatively short period of time ensures that it is
being mechanically aligned while still having over 50\% of
crosslinking to take place. Finally, while still constrained in
the clamps, the sample was placed in an oven at 70$^{\circ}$C for
a further 24 hours as it finished its crosslinking cycle under
stress. As a result a homogeneous elastomer was prepared where the
nanotubes had a preferred orientation induced by the processing
technique and are also well dispersed in the matrix.  The degree
of nanotube alignment in each sample was quantified using X-ray
techniques (discussed below).

There is a separate question of solvent and shearing conditions,
and the time required for the full MWCNT dispersion; systematic
studies of nanotube dispersion and re-aggregation rates are to be
published shortly. The quality of nanotube dispersion is monitored
throughout the processing with the use, initially, of optical
microscopes and later with a High-Resolution Scanning Electron
Microscope (HRSEM, Phillips XL 30 series) as aggregate sizes
reduce below optical resolution. We find that a shearing regime of
high-viscosity mixture, lasting 24 hours, is suitable in removing
nanotube aggregates. Samples are identified by the wt\% of MWCNTs
mixed with the PDMS and the abbreviations assigned to them in
table~\ref{tableMats}. Most experiments have been conducted on the
0, 0.02, 0.3, 0.5, 1, 2, 3, 4 and 7wt\% MWCNTs in PDMS elastomer
films. A sample with 3wt\% carbon black instead of nanotubes has
also been made using the same procedure.

\subsubsection{Nematic elastomer composite preparation}

There is a wealth of literature regarding liquid crystal elastomer
(LCE) preparation~\cite{Eugenebook}. For our purposes, we have
tested two specific types of LCE: polydomain and monodomain, with
uniaxially aligned nematic director. Control samples containing no
nanotubes were made, following the procedure introduced
in~\cite{Kupfer1991LCE} and widely used in the field since. The
procedure of nanocomposite preparation was detailed
in~\cite{Courty2003}. The polysiloxane backbone chains ($\sim 60$
monomer units long) had their Si$-$H bonds reacted, using platinic
acid catalyst, with the terminal vinyl groups of the mesogenic
rod-like molecule $4-$methoxyphenyl) $-4'-$buteneoxy benzoate
(MBB) and the two-functional crosslinker
$1$,$4-$di-$11-$undeceneoxy benzene (11UB), with the molar ratio
18:1 (thus achieving the 9:1 ratio of substituted groups on each
chain, or the effective 10\% crosslinking density). The
crosslinking was initiated by a combination of adding the catalyst
and heating to 80${}^{\rm o}$C in the already described
centrifugation reaction chamber. The subsequent procedure of
two-stage crosslinking, with intermediate stretching to induce
director alignment, is similar to the procedure of PDMS alignment
above.

Polydomain control samples were made identically with the single
exception that no uniaxial extension is applied during the
crosslinking cycle. This avoids orientational bias being
introduced during processing.

For LCE nanocomposites, a minor modification is made. Before the
crosslinker and catalyst were added, MWCNTs were shear mixed into
the polymer to ensure homogenous dispersion. Due to the
sensitivity of the crosslinker and catalyst, shear mixing is
reduced to 4 hours at elevated temperatures ($\sim$50$^{\circ}C$.
This is acceptable as the nanotube concentration in such systems
was very small (0.15-0.2wt\%) while the nematic polymer is highly
viscous. A higher concentration in such a system is currently
unachievable due to catalyst and crosslinker sensitivity
limitations.

\subsubsection{SIS composite preparation}

The SIS nanocomposite was made by adding the desired quantity of
tubes (0.01wt\%) to the melt of SIS symmetric triblock copolymer
(14\% of polystyrene, obtained from Sigma Aldrich) in the presence
of small amount of toluene solvent. The solvent dilutes the
otherwise rubbery thermoplastic system and allows shear mixing at
40$^{\circ}C$ for 24 hours. The solvent was added in small
portions during the mixing cycle to maintain the mixture in a
high-viscosity state. Once the dispersed state was achieved,
fibers could be drawn from the mixture and left to air dry. During
this period the PS micelles are formed in the usual way
\cite{SISbook} to form the elastic network surrounding and
encapsulating the nanotubes. We note that too high a loading of
MWCNT prevents physical crosslinks from occurring in the host
polymer and thus nanotube content was kept to a very low level.

In all cases the sample dimensions were kept approximately
constant, 1.5mm $\times$ 3cm, with thickness 0.2mm.

\subsection{Experimental techniques} \label{expt}

\begin{figure}
\centering \resizebox{0.22\textwidth}{!}{\includegraphics{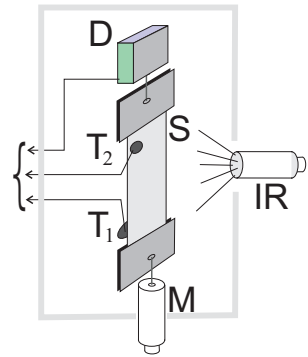}}
\caption{Scheme of the apparatus; see text for detail. }
\label{appa}
\end{figure}

The main part of this study, and purpose of this paper, is
concerned with the response of these materials to infrared
radiation and to that end a specially constructed rig was built to
test the actuator response. Two dynamometers were used in this
study; a 25g dynamometer for small sensitive measurements and a
larger 55g dynamometer allowing a larger range of responses to be
tested. The dynamometers (Pioden Systems Ltd) were housed in a
custom made thermal-control box with an open front end. The
device, together with an independent thermocouple, outputs data
via a DAQ card to a PC, see Fig.~\ref{appa}. The sample (S) was
clamped in the frame with its length controlled by the micrometer
(M), with $\pm$0.001mm accuracy, and the exerted force measured by
the dynamometer (D). Thermocouples ($T_1$ and $T_2$) were placed
in front and behind, on the sample surface. The actuation was
induced by the light source (IR), Schott KL1500 LCD, with quoted
peak power density at $\approx$675nm, 702 $\mu$W/cm$^{2}$ at 1m
distance. The source uniformly illuminated the sample from $\sim
2$cm distance. Measuring the scaling of the intensity decay with
distance, we obtained that the power density delivered to the
sample was $\sim 1.5 \hbox{mW/cm}^2$ at 675nm. The rig was
enclosed in the thermally controlled compartment, and calibrated
with weights to give a direct measure of stress and strain.

\begin{figure}
\centering \resizebox{0.35\textwidth}{!}{\includegraphics{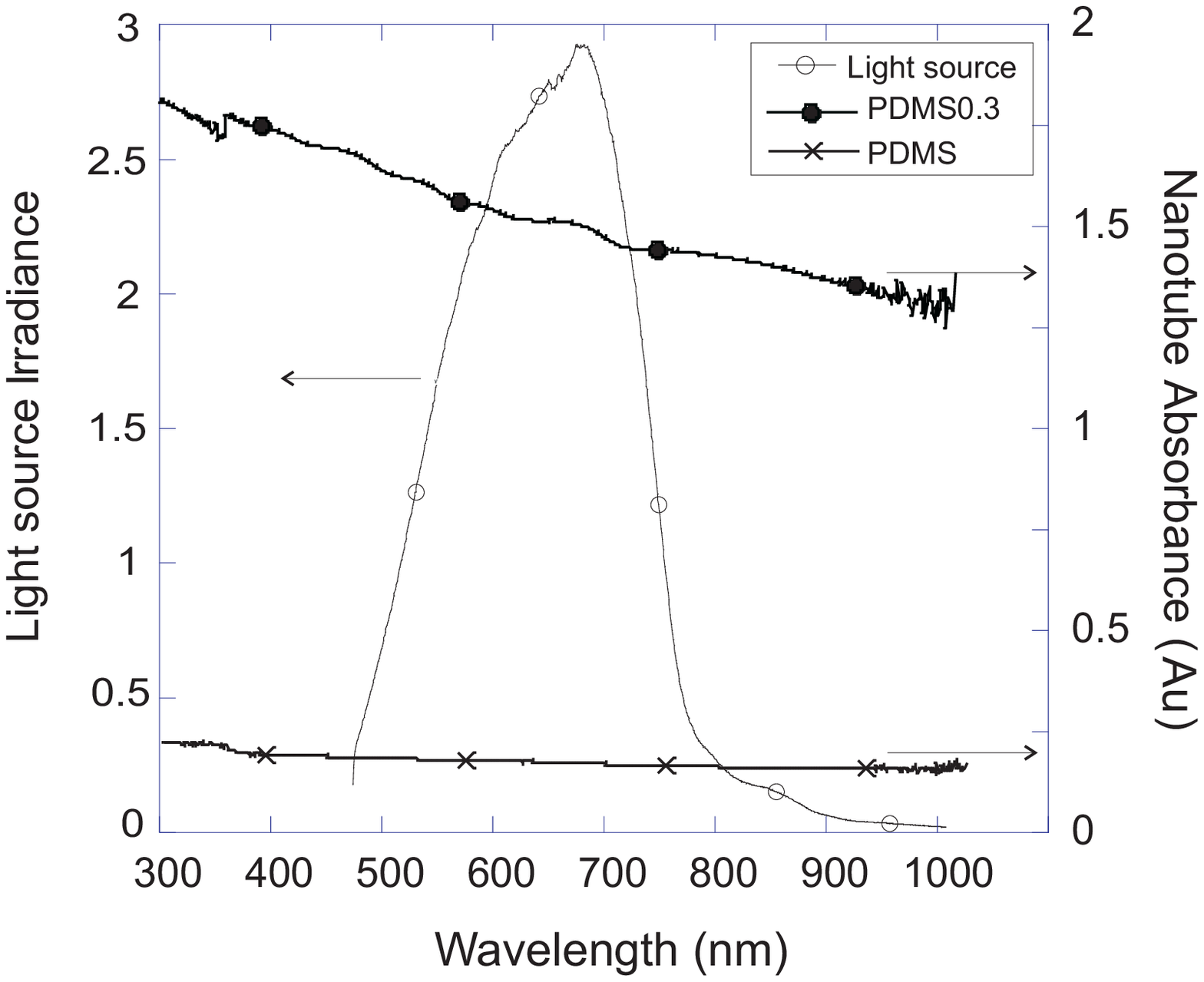}}
\caption{Spectral data of the light source (left axis, arb.
units), and the normalized absorbtion of the PDMS0.3 composite and
the control pristine PDMS elastomer (right axis). }
\label{Absorbance}
\end{figure}

Figure~\ref{Absorbance} shows the spectral distribution of the
light source, as well as the nanotube absorbance. These
measurements were carried out on a Varian Cary 300 BIO UV-visible
spectrophotometer in the 190-1000nm range, adjusted for the
background. Absorbtion units {\sf Au}$=\log [I_0/I_{\rm
transmitted}]$ indicate that the  PDMS control sample of given
thickness transmits $\sim 70$\% of light across the spectrum. In
contrast, the same thickness of low-loading PDMS0.3 composite
absorbs $> 97$\% of light across a range of wavelengths. The
strong absorbtion of light by nanotubes is a well-known effect,
although the relatively flat spectral distribution was a surprise
in our case, Fig.~\ref{Absorbance}.

To standardize the results across all samples, pre-experimental
checks were undertaken to accurately find the zero strain value of
each experiment. The gradient of the stress-strain curve for a
buckled sample was equated with the gradient for the stress-strain
curve of the taut sample -- the meeting point of the two lines
designates the zero-point strain, with the length of sample
defined as $L_0$. The imposed extensional strain is calculated by
$\varepsilon= (L-L_0)/L_0$, with $L$ provided from the micrometer
reading.

After a fixed pre-strain was applied to each sample, the stress
was allowed to relax for a minimum of 10 minutes. After this
relaxation period, readings of stress were taken for 1-2 minutes,
to verify that the material is equilibrated, and then the IR
source was switched on to full intensity. After a period of
exposure, the light source was switched off and further relaxation
data collected. After completion, the sample was relaxed and then
this protocol was repeated for a different applied pre-strain
$\varepsilon$. Each sample is tested under a range of applied
pre-strains between 2\% and 40\% ($0.02 \leq \varepsilon \leq
0.4)$. In order to avoid a systematic influence of pre-strain,
through thermal history and possible degradation, we applied the
different values of $\varepsilon$ in random order, not
sequentially. The LCE composites have been tested for even larger
deformation as they can spontaneously undergo thermal strains of
hundreds of percent~\cite{Eugenebook}.

\begin{figure}
\centering \resizebox{0.44\textwidth}{!}{\includegraphics{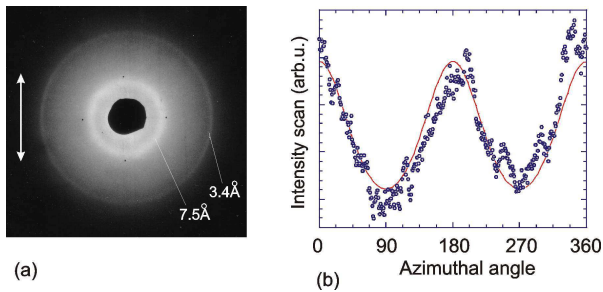}}
\caption{(a) \ The X-ray scattering image showing key reflections;
the outer ring (3.4${\AA}$) is the signal from the multiwall
nanotubes. The inner ring (7.5${\AA}$) represents the PDMS mesh
size, see section~\ref{xrayData}. The arrow shows the direction of
the uniaxial aligning strain. \ (b) \ The typical azimuthal
intensity variation, $I(\beta)$, at a scattering angle of
3.4${\AA}$ reflection. The data is fitted by the model
\cite{Deutsch1991}.} \label{xray}
\end{figure}

Our attempts to rationalize the observed response, changing
qualitatively on increasing the applied pre-strain, invoke the
concept of increasing nanotube alignment under uniaxial
deformation. To monitor this, wide angle X-ray diffraction
measurements were carried out on a Phillips PW1830 Wide Angle
x-ray generator (WAXS) using Cu$_{K_{\alpha1}}$ radiation (1.54$
\AA $), running at 40kV and 40mA. A specially designed clamp was
used allowing measurement of the X-ray images as a function of the
applied strain during the experiment. Azimuthal scans of intensity
were generated, Fig.~\ref{xray}, and fitted with theoretical
models. The method of background correction employed is crucial.
Two to three different regions were selected from an image to
gather an average of background noise which is then subtracted
from the azimuthal curves generated. This is repeated for all
scattering images before order parameter was calculated.

With the IR irradiation, the question always exists, whether the
response is due to photon absorption, or the trivial heating of
the materials (which does take place during irradiation). The
technique used to measure temperature involves two thermocouples
and we were reasonably sure that the measured increase in
temperature ($\sim$15-20${}^{\circ}$C) is a true temperature
across the sample. A separate study was conducted using
thermocouples on the surface and embedded within the sample which
showed similar values throughout for any relevant time-scale. The
samples were kept purposely thin to ensure very quick heat
conduction. To compare the effects, the same experiment was
carried out on the PDMS1 sample, with the infrared source replaced
by a mica-insulated heater (Minco Products Inc.) mounted
approximately 10mm away from the sample. Temperature was regulated
through an integrating controller using thermocouples mounted on
the sample. The maximum temperature reached was 15-20$^\circ$
above ambient, and although thermo-mechanical response was
present, it was much slower and almost an order of magnitude
smaller than the direct IR-irradiation effect.

\section{Material characterization} \label{mat-sec}
\subsection{Elastic strength}

\begin{figure} 
\centering \resizebox{0.3\textwidth}{!}{\includegraphics{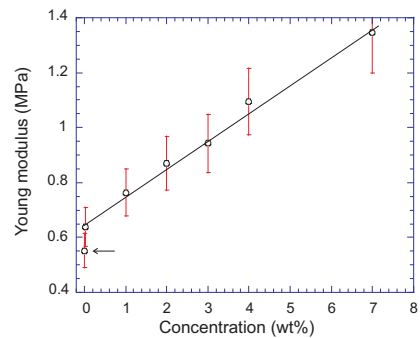}}
\caption{Young modulus $Y$ for PDMS nanocomposites at increasing
MWCNT loading. The arrow points at the value for control PDMS
rubber. } \label{modu}
\end{figure}

Figure~\ref{modu} shows a summary of the linear mechanical
response of our nanocomposites for different nanotube loadings in
the crosslinked PDMS matrix. As the concentration of MWCNTs is
increased the rubbery network becomes stiffer and Young modulus
$Y$ (the response to static linear extension) of the composites
increases. This is expected and in line with literature
findings~\cite{Gorga2004MechAlign,Harris2004}. An account for
subtle variations in measured moduli could be obtained from the
analysis of the polymer-nanotube interface and relaxation of local
stress in the composites. This is not the focus of the work
presented here.

At very low nanotube loading one might expect that large regions
of rubbery network are still pristine. However, even with the
lowest nanotube loading ($\sim$0.02wt\%), a small but significant
increase in modulus was observed suggesting that the presence of
the tubes even in tiny quantities has an immediate mechanical
effect. Starting from $\sim$0.02wt\% the linear increase of the
modulus was observed, characteristic of non-interacting inclusions
in the elastic matrix. We observe an almost three-fold linear
increase in the elastic modulus from 0-7wt\%.

It can be argued that at higher loading the concentration
dependence must become non-linear, quadratic at first indicating
the pair interactions between nanotube inclusions, etc. This may
signify the onset of a `mechanical' percolation within the
composite system. One may be tempted to make a connection between
the onset of this non-linear regime and the separately determined
electric percolation threshold, when the composite becomes
conducting through nanotube contacts. Unfortunately, it is
difficult to make an unambiguous connection in a crosslinked
system: an increase in tube concentration would undoubtedly
increase the modulus $Y$ -- but would also cause a reduction in
the crosslinker concentration (the presence of nanotubes has an
inhibitive effect on siloxane reactions). Overall, the Young's
modulus of such a nanocomposite would not be able to directly
reflect the nanotube interactions.

For completeness, let us quote the measured Young modulus values
for the other nanocomposite systems under study -- MLCE0.2:
$Y\approx 0.2$~MPa, PLCE0.15: $Y\approx 0.2$~MPa, SIS0.01:
$Y\approx 0.6$~MPa.

\subsection{Nanotube orientation by strain}\label{xrayData}
\subsubsection{X-ray data analysis}
 As a crucial part of
material characterization, before and during the main actuation
experiment, we need a more quantitative analysis of the nanotube
orientation in the matrix. It is a key element in our model of the
actuation mechanism, but also has its own merit considering the
high interest in all aspects of polymer nanocomposite studies.
Wide angle X-ray diffraction is used as a method to determine the
average tube orientation as a function of increasing applied
uniaxial strain. Figure~\ref{xray}(a) shows characteristic
features of the diffraction image. The image is for PDMS7,
initially non-aligned, stretched by $\varepsilon = 0.33$ (33\%).
The scattering reflection at an angle corresponding to MWCNT [002]
layer periodicity (inter-shell spacing~\cite{Charlier1993}) of
3.4$\AA$ allows calculation of the tube orientation distribution
from the corresponding azimuthal intensity variation,
Fig.~\ref{xray}(b).

In this separate study of deformation-induced alignment we used a
7wt\% loaded PDMS7 composite simply to enhance the X-ray contrast
and enable using a desktop X-ray generator (as opposed to the
synchrotron study required for very low-loading composites).  Note
that the scattering intensity at 3.4$\AA$ is still relatively low,
because of the small contrast between the nanotubes and PDMS
matrix.

A question must arise about the bright scattering ring
corresponding to the length scale $\sim 7.5\AA$. This is a very
interesting feature, but totally irrelevant for our work: this
scattering is exactly the same in the pristine PDMS rubber
prepared in the same batch. In the PDMS network, with no solvent,
the only X-ray contrast may arise due to the difference between
the chains and crosslinks. A very clear scattering length must be
an indication of crosslink density fluctuations (in other
terminology called clustering). As the extensive theory of this
clustering phenomenon suggests~\cite{rabin}, at the given chain
lengths and crosslinking density the network is well below the
`crosslink saturation threshold' and the correlation length of
clustering should be of the order of network mesh size. The length
scale of $\sim 7.5\AA$ is very accurately this size and,
accordingly, we believe this scattering to be produced by very
small scale crosslink density fluctuations. These should not
affect macroscopic properties, or even the local MWCNT embedding.

Intensity variation along the azimuthal arcs, $I(\beta)$ in
Fig.~\ref{xray}(b), is the signature of the orientational
distribution function. When $I(\beta)$ is approximated as a
Legendre polynomial series in $\cos \beta$, it gives a measure of
the orientational order parameter S$_d$:
\begin{equation}\label{orientation}
S_d \equiv \langle P_2 \rangle
={\textstyle{\frac{3}{2}}}(\langle\cos^2\beta \rangle-1),
\end{equation}
where the averaging is performed with $I(\beta)$ as the
distribution function. This is called the Herman's orientation
parameter and it adequately describes the true orientational
ordering at very small bias, when $S_d \ll 1$.

At higher degree of alignment (such as, for instance, in nematic
liquid crystals) the orientational distribution function
significantly deviates from the measured $I(\beta)$. The analytic
treatment of the problem of X-ray scattering from orientationally
biased medium is developed by Deutsch~\cite{Deutsch1991}, mainly
in the context of nematic liquid crystals. Instead of using the
full theory, we have derived an interpolating analytical
approximation to the complete results of~\cite{Deutsch1991}. With
that, the orientational order parameter is given by
\begin{eqnarray}\label{DeutschOrientation}
S_d &=& 1-\frac{3}{2N} \int_{0}^{\pi/2}I(\beta) \sin \beta \bigg\{
 \sin\beta \\
 && \qquad \qquad + \cos^{2}\beta \ln
\left[\frac{1+\sin\beta}{\cos\beta} \right] \bigg\} d\beta
\nonumber
 \end{eqnarray}
$${\rm with} \qquad N=\int_{0}^{\pi/2}I(\beta)d\beta  .$$
This expression also properly accounts for nontrivial geometric
factors involved in projecting the 3D orientational distribution
onto a 2D detector plane. Experimental data was analyzed using
both Herman's approximation and Deutsch's interpolated analytic
result. We conclude that in the range of parameters we are working
with both expressions were in agreement qualitatively but slightly
differ quantitatively. We favor the Deutsch analytical method and
used it exclusively in this study.

\subsubsection{Induced orientation of nanotubes} \label{orie}

\begin{figure}
\centering \resizebox{0.32\textwidth}{!}{\includegraphics{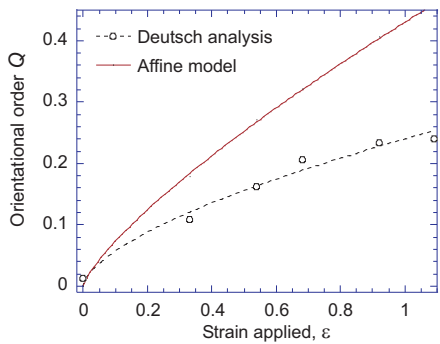}}
\caption{The change in the orientational order parameter $S_d$ of
nanotubes in PDMS7 composite, as a function of imposed uniaxial
strain, obtained from the X-ray scattering data ($\circ$ - data
points; dashed line is a guide to the eye). Solid line shows the
affine rigid rod model prediction. } \label{order}
\end{figure}

Figure~\ref{order} presents the results of the calculation of
orientational order parameter $S_d$, acquired as a function of
sample strain applied to the PDMS7 sample, as well as the
prediction of the theoretical model discussed below. As the
applied strain is increased, the initially disordered nanotubes
align along the strain axis resulting in bias in the azimuthal
curve $I(\beta)$. This phenomenon has recently been confirmed by
synchrotron experiments~\cite{Kelarakis2005} although it should be
noted that the focus of the work by Kelarakis \textit{et al.} was
not on nanotube reorientation in a rubbery matrix. Our composites,
with no significant initial alignment, on subsequent stretching
reached substantial values of induced orientational order.
Furthermore, the change in orientation on stretching was
reversible, i.e. equilibrium, which is discussed later. To our
knowledge, this is the first time nanotube \emph{reorientation}
has been reported and analyzed in a semisolid/rubbery sample.

As will be described in section~\ref{ir-sec}, there is good
evidence that much better nanotube alignment can be achieved if
dispersed in a monodomain liquid crystal elastomer during
processing -- the mesogenic moieties act to align the tubes. A
similar effect has been demonstrated for pure liquid
crystals~\cite{Lynch02,Dierking2004LC}, and also is well-known in
the field of ferronematics~\cite{ferronematic}. X-ray diffraction
of such a system is not reported due to the continuing problem of
poor contrast between the two species and only low nanotube
concentrations studied.

There is an issue, well argued in the
literature~\cite{Somoza2001,Islam2004}, about whether a truly
isotropic nanotube dispersion can be obtained. Regarding the tubes
as rigid rods with extremely high aspect ratio, well dispersed in
an amorphous medium, the Onsager transition to the steric
orientational ordering could start at very low concentrations as
has been recently reported~\cite{Song2005}. We have as yet
observed no clear indication of truly nematic liquid crystalline
architecture in our system, although this could be due to a number
of factors including matrix viscosity and sample preparation.

\subsubsection{Affine model of induced orientation} \label{affQ}

\begin{figure} 
\centering \resizebox{0.3\textwidth}{!}{\includegraphics{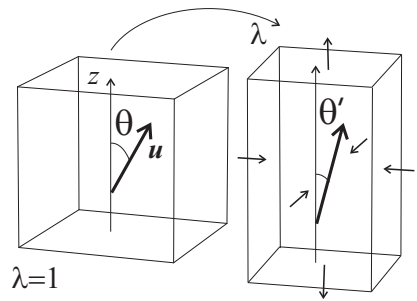}}
\caption{The scheme of an affine incompressible extension,
changing the orientation of an inflexible rod embedded in the
medium.} \label{defo}
\end{figure}

Let us compare the observed induced orientational order parameter
$Q(\varepsilon)$ with a simple model prediction based on the
affine deformation of the rubbery matrix. The most straightforward
approach is to evaluate the average orientational bias resulting
from an imposed uniaxial extension of such a matrix, in which the
ensemble of rigid rods is initially embedded isotropically. The
direction, known as the director $\bm{n}$, is the average axis
along which nanotubes can and do align. This is a local property
of the system obtained as a result of averaging of individual
particle axes, $\bm{u}_i$, over the macroscopically infinitesimal
volume. This averaging applies equally well for rigid rod-like
particles and for the segments of semi-flexible chains, e.g. in
the study of nematic polymers \cite{Eugenebook}. The corresponding
local orientational order parameter is a second-rank tensor
$Q_{\alpha \beta}$ which for the uniaxial alignment (reflecting
the quadrupolar symmetry breaking) is defined as:
\begin{equation}\label{orientation0}
Q_{\alpha \beta} \equiv {\textstyle{\frac{3}{2}}} Q (n_\alpha
n_\beta - {\textstyle{\frac{1}{3}}} \delta_{\alpha \beta} )
 \equiv \left(
\begin{array}{ccc} -\frac{1}{2}Q & 0 & 0 \cr 0 &-\frac{1}{2}Q & 0
\cr 0 & 0 & Q
\end{array} \right),
\end{equation}
where the principal axes are aligned with $z$ along the uniform
ordering direction $\bm{n}$, cf. Fig.~\ref{defo}. The value of the
local scalar order parameter is indeed the average of the second
Legendre polynomial of orientation of embedded rods,
\begin{equation}
Q \equiv S_d  = \int_{0}^{\pi} [{\textstyle{\frac{3}{2}}} \cos^2
\theta -{\textstyle{\frac{1}{2}}}] \, P(\theta) \, \sin\theta
d\theta d\varphi . \label{defQ}
\end{equation}
Here $(\bm{n}\cdot \bm{u}_i) \equiv \cos \theta_i$ for each rod,
and $ P(\theta)$ is the orientational probability distribution,
normalized such that $\int P(\theta) \sin \theta d\theta d\varphi
= 1$.  Let us assume the initial state is un-aligned, and thus
characterized by the flat distribution $P_0(\theta) = 1/(4\pi)$.

The uniaxial extension of an incompressible elastic body is
described by the matrix of strain tensor
\begin{equation}
\bm{\Lambda} = \left(
\begin{array}{ccc} 1/\sqrt{\lambda} & 0 &
0 \cr 0 & 1/\sqrt{\lambda} & 0 \cr 0 & 0 & \lambda
\end{array}  \right)  , \label{lambda}
\end{equation}
where the axis of stretching is taken as  $z$ and the magnitude of
stretching is $\lambda = 1+\varepsilon \equiv L/L_0$ is the ratio
of the stretched and the initial sample length along $z$,
Fig.~\ref{defo}. This tensor describes the affine change of shape,
which could also be visualized as locally transforming an embedded
sphere (representing the orientational distribution $P_0$) into
the ellipsoid (representing the induced orientational bias) of the
same volume and the aspect ratio $R_\| / R_\bot = \lambda^{3/2}$.

After such a deformation, every element of length in the body
changes affinely according to the matrix product
$\bm{L}'=\bm{\Lambda}\cdot \bm{L}$, which in our case of uniaxial
incompressible extension means that $L_z'=\lambda L_z$ and
$L_\bot'= (1/\sqrt{\lambda}) L_\bot$. This corresponds to the new
angle of the rod, $\theta'$ such that $\tan \theta' =
L_\bot'/L_z'=(1/\lambda^{3/2}) \tan \theta$. Therefore, to obtain
the new (now biased) orientational distribution function we need
to convert the variable $\theta$ into the new (current) variable
$\theta'$, which gives (after some algebraic manipulation)
 \begin{eqnarray}
 \theta
& \rightarrow & \arctan (\lambda^{3/2} \tan \theta'); \nonumber \\
 \sin \theta \, d\theta & \rightarrow & \frac{\lambda^3}{(\cos^2
\theta' + \lambda^3 \sin^2 \theta')^{3/2}}\sin \theta' d\theta' .
 \end{eqnarray}
This defines the expression for the normalized orientational
distribution function
\begin{equation}
P(\theta')=\frac{\lambda^3}{4\pi (\cos^2 \theta' + \lambda^3
\sin^2 \theta')^{3/2}} , \label{ptheta}
\end{equation}
which is an explicit function of the uniaxial strain applied to
the body and can be used to calculate the induced order parameter
$Q$:
\begin{equation}
Q(\varepsilon) = \frac{3}{2} \int \frac{\cos^2 \theta'
[1+\varepsilon]^3 \sin \theta' d\theta'd\varphi}{4\pi (\cos^2
\theta' + [1+\varepsilon]^3 \sin^2 \theta')^{3/2}} - \frac{1}{2}.
\label{orderaffine}
\end{equation}
Analytical integration of this expression gives a function
$Q(\varepsilon)$, which is plotted as a solid line in
Fig.~\ref{order}:
\begin{eqnarray}
Q(\varepsilon) &=& \frac{3+2\varepsilon
(3+3\varepsilon+\varepsilon^2)}{2\varepsilon
(3+3\varepsilon+\varepsilon^2)} \label{orderfull}\nonumber \\
&& + \frac{3(1+2\varepsilon
(3+3\varepsilon+\varepsilon^2))^3}{4\varepsilon
(3+3\varepsilon+\varepsilon^2)\sqrt{1-(1+\varepsilon)^3}}\ln
B(\varepsilon)\nonumber
\end{eqnarray}
$ \rm{where} \ B(\varepsilon)=\left[
\frac{-1+(1+\varepsilon)^3+\sqrt{1-(1+\varepsilon)^3}}
{1-(1+\varepsilon)^3+\sqrt{1-(1+\varepsilon)^3}} \right].$

\noindent At relatively small strains, it approaches the linear
regime: $Q \approx \frac{3}{5}\varepsilon - \frac{6}{35}
\varepsilon^2+ ...$.

The experimental data displays a lower order parameter than that
predicted by the affine model, although has the same qualitative
trend. One must remember that the model presented here does not
account for tube flexibility. Also, some proportion of the tubes
would be unable to orientate affinely due to the entanglements.
The experimental data reflects this and, accordingly, gives
slightly lower values of order parameter.

\section{Infrared Actuation} \label{ir-sec}

\subsection{Typical observations}

\begin{figure} 
\centering \resizebox{0.4\textwidth}{!}{\includegraphics{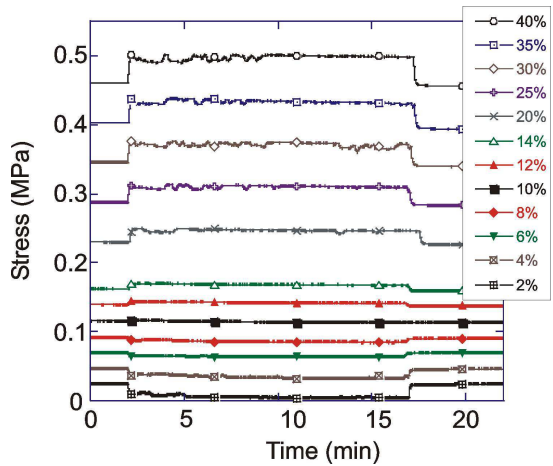}}
\caption{The response of a 1wt\% nanocomposite PDMS1 to IR
radiation at different levels of pre-strain $\varepsilon$. Stress
is measured at fixed sample length (different pre-strain curves
labelled on the plot).} \label{steps}
\end{figure}

The detailed response to infrared stimuli is presented in
Figs.~\ref{steps}, showing the stress measured in the PDMS1
sample. Results for all composites are qualitatively similar. We
shall later examine the dependence on the host polymer and the
tube concentration. Composites, initially un-aligned, are
subjected to an increasing extension that we call pre-strain
$\varepsilon$. At each $\varepsilon$, the IR-irradiation takes
place and the stress response recorded. The complexity of the
plots necessitates more detailed description of what takes place.

We begin with a 2\% pre-strain ($\varepsilon=0.02$) applied to it
initially. At $t=2$mins the light source is switched on and the
stress reading changes downwards, meaning that the sample natural
length $L_0$ has expanded on actuation (recall that the actual
length $L$ is fixed through $\varepsilon=L/L_0-1$). After a period
of constant irradiation, at $t=15$mins, the light source is
switched off -- and the stress reading returns to its original
value. This experiment is then repeated with the same sample
pre-strained at different values, up to 40\%, as shown by the
sequence of stress-reading curves in Fig.~\ref{steps}.

\begin{figure} 
\centering \resizebox{0.36\textwidth}{!}{\includegraphics{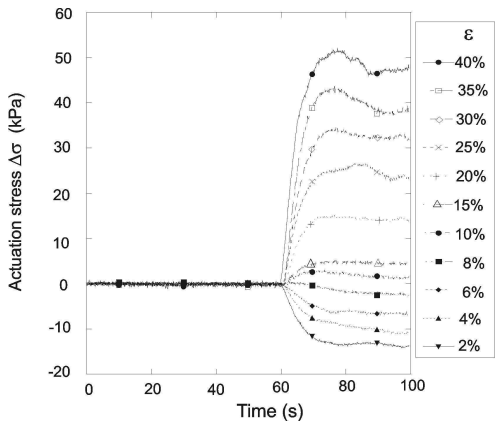}}
\caption{The speed of actuation response, illustrated by plotting
the actuation stress in PDMS3 nanocomposite, $\Delta \sigma$ in
kPa, as a function of time for different pre-strain values
(labelled on the plot). } \label{speed}
\end{figure}

\begin{figure} 
\centering \resizebox{0.37\textwidth}{!}{\includegraphics{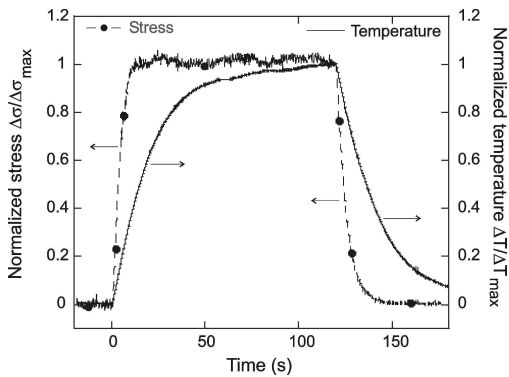}}
\caption{ The normalized stress response plotted alongside the
normalized change in temperature, as functions of time (PDMS3,
pre-strain $\varepsilon = 20\%$), see text for discussion.}
\label{steps2wt}
\end{figure}

The data in Fig.~\ref{speed} is assembled to demonstrate the speed
of the actuation process more clearly, while Fig.~\ref{steps2wt}
helps differentiate between the light- and heat-driven actuation
response. In this case the data is for a PDMS3 composite; as was
mentioned above, all materials exhibit the same qualitative
features. We plot the change in stress and change in temperature,
normalized by their maximal value at saturation in the given
experiment; plotted in this form, all the results (for different
tube loading and different pre-strain) appear universal.

The change in temperature by IR-heating is unavoidable and reaches
$\Delta T \sim 15^\circ$C maximally on the sample surface, in our
setting (thermocouples placed below the surface and embedded in
the center of the sample may report the temperature change of up
to $20^\circ$C depending on nanotube concentration, but we avoided
disturbing the sample in mechanical experiments). This highlights
an important question as to whether the mechanical response is due
to the photon absorption or plain heat. Figure~\ref{steps2wt}
shows that the stress reaches its peak and saturation in $\sim
0.5$min, while the thermal takes over 2min to reach its peak.
Although the difference in rates is not very dramatic, the fact
that the stress response is faster suggests that its mechanism is
not caused by the trivial sample heating. In a separate study (not
shown) we reach the conclusion that thermo-mechanical effects do
exist (i.e. the MWCNT-loaded composite has a stronger mechanical
response to heating than a pristine polymer) but their magnitude
is almost an order of magnitude smaller than the direct IR-photon
absorption mechanism.

\subsection{Analysis of IR-actuation response}

Of great interest is the observation that this response changes
sign at a certain level of pre-strain (at $\varepsilon \sim 10\%$
in Fig.~\ref{steps}). In other words, relaxed or weakly stretched
composites show the reversible \textit{expansion} on irradiation,
while the same sample, once strained more significantly,
demonstrates an increasing tendency to \textit{contract} (hence
the increase in the measured stress). This is our key finding.

\begin{figure} 
\centering \resizebox{0.4\textwidth}{!}{\includegraphics{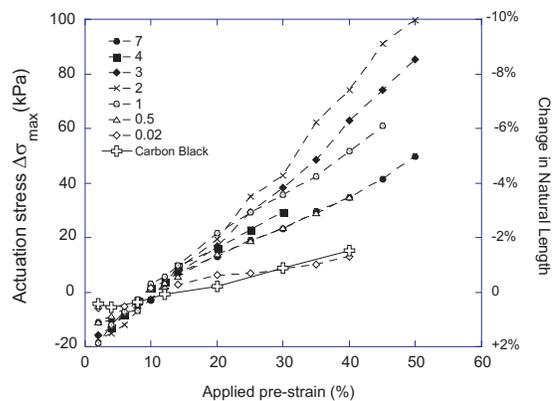}}
\caption{The magnitude (in kPa) of exerted actuation stress (the
height of steps in Fig.~\ref{steps}, $\Delta \sigma_{\sf max}$),
as function of pre-strain. Different PDMS composites are labelled
on the plot by their wt\% value. The right $y$-axis shows the
corresponding actuation stroke: the change in natural length
$L_0$(IR).} \label{sum}
\end{figure}

Figure~\ref{sum} summarizes the magnitude of the IR-actuation
effect by plotting the stress step at saturation ($\Delta
\sigma_{\sf max}$) in the IR-on state, at different levels of
pre-strain and for samples with increasing MWCNT loading. Although
this is not explicitly measured in our (isostrain, $L=$const)
experiment, we can directly calculate the change of the underlying
natural length $L_0$(IR) of the samples on actuation from the
known Young modulus values. This is shown on the right axis of the
same plot, highlighting the regions of expansion and contraction.
Remarkably, all samples with different nanotube loading appear to
have a crossover at the same point, around 10\% pre-strain.

For comparison, the pristine PDMS rubber in the same experiment,
shows no discernible stress response at all. Also, the response of
the PDMS composite with a 3wt\% of carbon black is much lower.
Indeed, this 3wt\% carbon-black composite closely follows the
low-concentration PDMS0.02 composite. We believe the response is
due to trace amounts of nanotubes that can often be found in
commercially supplied carbon black. Hence the very small response
from such a highly loaded sample. The shift in transition
pre-strain may well be due to the trace nanotubes having their
alignment hindered by the activated carbon black.

\begin{figure} 
\centering \resizebox{0.33\textwidth}{!}{\includegraphics{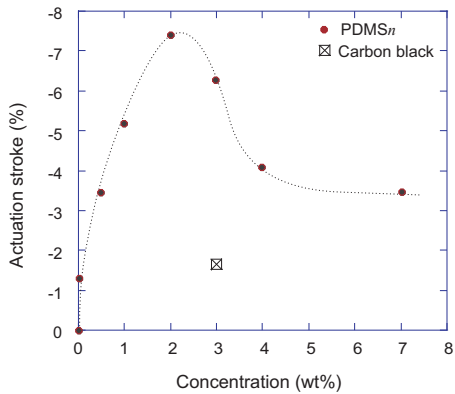}}
\caption{The magnitude of the actuation stroke at
$\varepsilon=40\%$ as a function of filler concentration $n$. The
maximum of the response at $\sim$2wt\% is evident. The single
square symbol gives the value for 3wt\% carbon black filler in
PDMS. } \label{Interplay}
\end{figure}

The interaction between filler particles is also evident when the
nanotube concentration is increased beyond 2wt\% loading. Above
this value, the magnitude of the actuation stroke decreases
sharply. Figure~\ref{Interplay} displays the effect clearly by
plotting the maximal change in natural length $L_0$ on IR
irradiation, at a fixed $\varepsilon=40\%$, for all PDMS-nanotube
composites, and the 3wt\% carbon black system for comparison. A
rapid increase in the stroke is observed with increasing
concentration, which then peaks at 2wt\% nanotube loading. The
reason for the subsequent decline is not obvious. There may well
be a number of factors that interplay to reduce the stroke
magnitude. At high concentrations entanglements between the long
tubes could take place. Note that through conservation of volume,
a contraction in the $z$-axis of the tube will be concomitant with
an expansion in the $x$-$y$ plane; such expansion may be hindered
for a significant number of nanotubes by their nearest neighbors
(another representation of entanglement). There may also be an
issue of photon screening at higher concentration which is
difficult to avoid.

\subsection{Observations in other host polymers}\label{hosts}

\begin{figure} 
\centering \resizebox{0.3\textwidth}{!}{\includegraphics{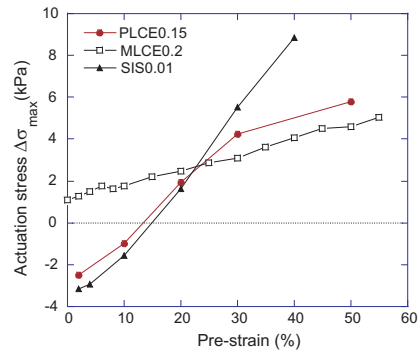}}
\caption{Summary of IR actuation stroke from LCE composites
(PLCE0.15 and MLCE0.2) and a SIS0.1 composite, as a function of
applied pre-strain. Note that MLCE data does not have a crossover
at $\varepsilon \sim 10$\%, since the tubes are aligned there at
preparation. For comparison, the similar data for two PDMS
samples, 0.5 and 0.02, is shown by dashed lines.}
\label{PolydomainCNT}
\end{figure}

Other polymers acting as a crosslinked host matrix for the
low-concentration nanocomposite display the same qualitative
behavior as PDMS systems. Figure~\ref{PolydomainCNT} summarizes
the response of LCE and SIS composites. The direction of the
actuation, changing from expansive to contractive mode with
increasing MWCNT alignment, as observed in PDMS-nanotube samples,
is unambiguously reproduced for vastly different materials.

The magnitude of the actuation stroke is shown in
Fig.~\ref{PolydomainCNT}, in comparison with some of the PDMS
composites. The value of actuation stress is different for various
polymeric systems considered in this work, which is due to the
different Young modulus (which we use to calculate the stroke from
the measured stress $\Delta \sigma$). We see that the stroke
magnitude in these differing materials is in the same range of
magnitudes. SIS0.01 has a much lower filler concentration, and
again its stroke is comparable to that of a similarly loaded
PDMS0.02. This important finding demonstrates the universality of
multi-walled nanotubes behaving as photo-actuators regardless of
the soft matrix they are in.

The response of nematic liquid crystal elastomers to heat is well
documented~\cite{Eugenebook}. Because they can be
thermo-responsive materials, the data in Fig.~\ref{PolydomainCNT}
is obtained by a complex procedure of subtraction of such
background effects. We do not go into its details, as this is
irrelevant to the main points of the present paper, however, must
emphasize that the plotted response highlights the effect of
nanotubes within the given matrix.

In Fig.~\ref{PolydomainCNT} we note that the MLCE0.2 sample shows
no expansive actuation and a crossover, while the similar
polydomain (un-aligned) composite does. This is clearly because
the two-step crosslinking at preparation of the monodomain
material involves aligning the mesogenic
groups~\cite{Kupfer1991LCE}. The embedded nanotubes align strongly
under such conditions, as others have found in ordinary liquid
nematics~\cite{Lynch02,Dierking2004LC}. As already discussed, the
expansive mode of actuation will only occur when the degree of
nanotube alignment is very low. It is important that the crossover
occurs at $\varepsilon^* \sim 10$\% for all studied materials with
nanotubes not aligned before pre-strain.

\section{Modelling the mechanism}
There are two main questions to answer: what mechanism is
responsible for such a large photo-mechanical response, and why
does it reverse its direction on sample extension?

We shall try to deduce the actuation behavior of individual tubes
from the macroscopic observations detailed above. We believe the
change of actuation direction on increasing sample extension is
due to the nanotube alignment induced by pre-strain, as described
in section~\ref{xrayData} and before. In the whole region of our
pre-strains, the orientational order induced in the MWCNT
distribution is, to a good approximation, a linear function of the
strain: $S_d \approx 0.6\varepsilon$ in the affine model. At the
crossover strain $\varepsilon^* \approx 0.1$, the value of the
order parameter would be $S_d^* \sim 0.06$. We now apply the same
ideas about the induced orientational bias and averaging of the
(hypothetical) individual nanotube response.

Let us assume this individual nanotube response to the IR photon
absorbtion is, in essence, a contraction -- because this is what
our data shows the better-aligned composite response to be. It is
easy to imagine why this could be for an initially rod-like tube:
on photon absorption it could generate instabilities in the form
of kinks, thus decreasing the net length due to the charge carrier
separation. The resulting elastic deformation would be most
pronounced in the already defect-dominated regions of the
nanotube. Such an explanation, based on concentration of induced
polarons~\cite{Verissimo2001,Perebeinos2005}, would also link with
the earlier observation of a similar actuation response under DC
electric field~\cite{Courty2003}. An alternative possibility is to
suggest that large (and fast) local tube heating~\cite{ignite}
causes the surrounding region of locally aligned elastomer to
contract and ``crush'' the nanotube. This version of microscopic
events does not contradict the discussion and the data in
Fig.~\ref{steps2wt}, which shows the (slow) global thermal effect.

\begin{figure} 
\centering \resizebox{0.2\textwidth}{!}{\includegraphics{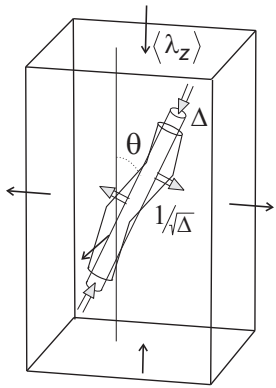}}
\caption{The scheme illustrating how the distortion (kinking) of
an individual tube, lying at an angle $\theta$ to the macroscopic
alignment axis, projects on the $z$-axis to contribute to the
average uniaxial strain, Eq.~(\ref{avL}). } \label{affi}
\end{figure}

At this stage we have to leave open the question of individual
tube response to near IR radiation. Using an affine approach
similar to the earlier analysis of ordering, let us assume that
each nanotube undergoes a linear contraction by a factor $\Delta =
R_\|({\rm IR}) /R_\|(0) < 1 $ (certainly proportional to radiation
intensity, which was kept constant in our work but has previously
been shown to effect the elastic
response~\cite{Zhang1999ElasticLight}). This contraction must be
accompanied by a transversely-isotropic volume conserving
expansion $1/\sqrt{\Delta}$. This means that a local strain is
created with the principal axes along the nanotube orientation (at
angle $\theta$ to the macroscopic $z$-axis, see Fig.~\ref{affi})
$$
\bm{\Lambda}_{\rm local} = \left(
\begin{array}{ccc} 1/\sqrt{\Delta} & 0 &
0 \cr 0 & 1/\sqrt{\Delta} & 0 \cr 0 & 0 & \Delta
\end{array}  \right) .
$$
The projection of this local strain on the macroscopic axis of
sample extension (and force measurement) is
\begin{equation}
\lambda_z({\rm IR}) = \Delta \cos^2 \theta + (1/\sqrt{\Delta})
\sin^2 \theta .  \label{lz}
\end{equation}
Averaging the local contribution with the probability to find the
nanotube at this orientation, $P(\theta)$ obtained in
section~\ref{xrayData}, gives an estimate of the effective stroke
of the actuation. When multiplied by the corresponding Young
modulus, the exerted stress of Fig.~\ref{sum} is also obtained:
$\Delta \sigma = Y (\langle \lambda_z \rangle -1)$ at small
deformations. Such a model is very crude indeed, ignoring a large
number of undoubtedly important and delicate factors of continuum
elasticity and nanotube response. However, it is elastically
self-consistent and has only one parameter, $\Delta$ that
presumably carries all the underlying complexity of the problem in
it.

The orientational averaging is given by using the distribution
$P(\theta)$ with the projection of local strain in Eq.~(\ref{lz}):
\begin{eqnarray}
\langle \lambda_z \rangle &=& \int_{0}^{\pi} [\Delta \cos^2 \theta
+ (1/\sqrt{\Delta}) \sin^2 \theta ] P(\theta) \, \sin\theta
d\theta d\varphi \nonumber \\
&\approx &  {\textstyle{\frac{1}{3}}} \left(\Delta
+2/\sqrt{\Delta} \right) -{\textstyle{\frac{2}{5}}} \varepsilon
\left( 1/\sqrt{\Delta} - \Delta \right) \label{avL}
\end{eqnarray}

Although the integral above has a full analytic form, it is more
transparent to present its expansion to the linear order of small
imposed pre-strain $\varepsilon$ as shown in the second line of
Eq.(\ref{avL}). This demonstrates the key point: at very low
pre-strain, $\varepsilon \rightarrow 0$, the average uniaxial
deformation of the disordered nanocomposite is positive
$(\lambda_z-1)$, i.e. the expansion of its natural length.
However, above a threshold pre-strain $\varepsilon^*$ this average
deformations transforms into the sample contraction along $z$. It
is easy to find the crossover,
\begin{eqnarray}
\varepsilon^* \approx
\frac{5(2-\Delta^{1/2}-\Delta)}{6(1+\Delta^{1/2}+\Delta)} ,
\label{estar}
\end{eqnarray}
so that the prediction would be to observe the crossover at
$\varepsilon^* \sim 0.1$ if the nanotube response factor $\Delta
\sim 0.8$. That is, on IR-irradiation the nanotube contracts
overall by $\sim$20\%. The value is higher than one might expect,
considering early reports in the literature of nanotube strains of
only 1-2\%. However, as Fig.~\ref{affi} indicates, our proposition
is not that of the lattice strain of nanotube walls but a
contortion of the tube as a whole. Although this has not been yet
directly observed and reported in the literature, a similar effect
of resonant undulation has been seen (in
simulation~\cite{YakobsonPRL1} and in
experiment~\cite{Poncharal1999}) in response to distortion beyond
the linear regime. Furthermore, the more recent theoretical work
on single-walled tubes supports the idea that massive $z-$axis
contraction. Although in our system the multi-walled tubes respond
under different conditions, being embedded in an elastic matrix
under strain and absorbing the IR photons, the overall distortion
factor of 20\% suggested by the model fit is perhaps not
altogether unreasonable.

\begin{figure} 
\centering \resizebox{0.4\textwidth}{!}{\includegraphics{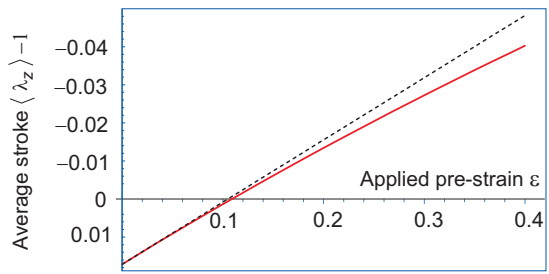}}
\caption{The result of the affine theoretical model,
Eq.~(\ref{avL}); the dashed line shows the linear approximation at
small pre-strain. Nanotube contraction factor is chosen to be
$\Delta = 0.8$, as suggested by the crossover strain value
$\varepsilon^*\sim 0.1$. } \label{theo}
\end{figure}

Figure~\ref{theo} plots the full (non-expanded) result of
orientational averaging of actuation stroke $(\langle \lambda_z
\rangle - 1)$ from the integral in Eq.(\ref{avL}) to illustrate
the points discussed in this section. The qualitative behavior (as
summarized in Fig.~\ref{sum}) is reproduced here almost exactly,
including the magnitude of the predicted actuation stroke [that
is, the ratio $L_0({\rm IR})/L_0(0)-1$].  Note that we use only
one parameter, $\Delta$ to match both the crossover
$\varepsilon^*$ and the actuation stroke magnitude, so the
conclusion is quite satisfactory and agrees with an apparent
universality discussed in section~\ref{hosts}. It is very likely
that the orientational feature of the effect, with its change of
actuation direction at a critical level of induced alignment, $S_d
\sim 0.06$, is captioned correctly, while much more work is
required to understand the individual nanotube response to IR
radiation generating the phenomenological factor $\Delta$ used in
this analysis.

\section{Conclusions}

To summarize, this work describes the rich photo-actuation
phenomena of carbon nanotubes embedded in crosslinked rubbery
matrices. The composite materials show the ability to change their
actuation direction, from expansive to contractive response, as
greater imposed strain is applied to the sample. We use differing
host polymers and confirm their relatively neutral role in the
actuation mechanism.

Theoretical models have been put forward to describe the
orientational order imposed on the nanotubes by a uniaxial strain
and the resulting actuation. Treating the nanotubes as rigid rods
that rotate affinely in a deforming matrix is a very simplistic
view, but it gives predictions that agree with experiment
qualitatively and often quantitatively. We believe that the
(certainly wrong) idea of the whole tube acting as a rigid rod is
not actually necessary -- in effect, in our model, the ``rigid
rods'' are nanotube segments below persistence length. In that
case, as in main-chain semiflexible nematic polymers, the model is
non-controversial and the agreement with experiment not
coincidental. The tube orientational distribution appears to
account well for the key macroscopic features of the observed
photo-actuation.

The strength of photo-mechanical response, at a given radiation
intensity, is of the order of tens of kPa. Translated into the
stroke, this corresponds to actuation strains of +2 (expansion) to
-10\% (contraction) depending on the nanotube concentration,
alignment (controlled by pre-applied strain) and the host matrix.
As expected, the response increases at higher nanotube loading --
however, only up to a limit. Beyond this limit ($\sim 2\%$ in
PDMS), the macroscopic actuation is inhibited by inter-tube
interactions and possible charge accumulation. The similar
(thermal actuation) behavior is also observed when the samples are
heated by the same amount, but this has a much lower amplitude.

Understanding the nature of the actuator mechanisms in this system
certainly warrants further theoretical and experimental
investigation. Many questions remain completely unclear, in
particular, what the effect would be if different types of
nanotube were used i.e. smaller multi-wall diameters, single-wall
tubes, various chirality etc. With actuating materials already
used in such widespread applications, from micromanipulators to
vibration control, the discovery of a structure that can respond
to stimulus in both directions may open new possibilities and
could mean an important new step toward finding applications for
nanotube based materials above and beyond improvements in existing
carbon fibre technologies.

\begin{acknowledgments}
We thank H. K\"orner, R. Vaia, A. Craig and P. Cicuta for useful
discussions, and K. Channon for help with nanotube absorbance
data. This work was carried out with the support of the
Engineering and Physical Sciences Research Council and a CASE
award from Makevale Ltd.
\end{acknowledgments}

\end{document}